%% file: main.tex
\documentclass{mc-abstract}

\usepackage{enumitem}

\year{2023}

\title{Back to the Future: \\From Microservice to Monolith}
\author{Ruoyu Su$^{1}$ \and Xiaozhou Li$^{1}$ \and Davide Taibi$^{1,2}$}
\affiliation{
    $^{1}$ University of Oulu, Oulu, Finland \\ 
    $^{2}$ Tampere University, Tampere, Finland 
}
\correspondence{ruoyu.su@student.oulu.fi; xiaozhou.li@oulu.fi; davide.taibi@oulu.fi}

\begin{document}

\maketitle

\abstract{Recently the trend of companies switching from microservice back to monolith has increased, leading to intense debate in the industry. We conduct a multivocal literature review, to investigate reasons for the phenomenon and key aspects to pay attention to during the switching back and analyze the opinions of other practitioners. The results pave the way for further research and provide guidance for industrial companies switching from microservice back to monolith.}

\section{Introduction}

Microservice has become an important style of architecture due to its decomposable and decentralized nature~\cite{3taibi2017microservices}. In recent years, microservice has become increasingly popular, especially in industry~\cite{1taibi2017processes}. Big companies like Netflix, Amazon and Spotify have also adopted microservice architecture and more and more companies are following this trend and migrating their systems to microservice~\cite{2taibi2018architectural}. They want to utilize the benefits of microservice, such as independent development, deployment and scaling to help the system solve the problem at hand, improve the quality of the system or facilitate software maintenance~\cite{5lenarduzzi2020does,4taibi2019monolithic}.

However, while several companies have made significant improvements in velocity and team independence, others did not achieve the benefits expected after migrating to microservice. With the increasing number of companies migrating from monolith to microservice, the drawbacks of microservice architectures are enhanced~\cite{soldani2018pains,taibi2018definition}.

Recently, there is a trend towards switching from microservice back to monolith. One example is Amazon Prime Video, which is one of the world's largest streaming services, serving millions of customers worldwide. It claims that the switch from a distributed microservice architecture to a monolithic application helps achieve greater scale, resilience and lower costs~\cite{PV1}. This report caused a heated debate among practitioners, after all, even big companies like Amazon have made rollbacks from microservice.

This paper aims to investigate the cases that switch from microservice back to the monolith with a multivocal literature review. Based on our goals, we define the following research questions: \textit{\textbf{RQ$_1$} What are the reasons for switching back to monolith? \textbf{RQ$_2$} What are the key aspects to pay attention to during the switching back? \textbf{RQ$_3$} What are the opinions of the other practitioners regarding such "switch-back"?}.

The results show that there are four cases that switch from microservice back to monolith of the company: Istio~\footnote{https://istio.io/}, Amazon Prime Video\footnote{https://www.primevideo.com/offers/nonprimehomepage/ref=dv\_web\_force\_root}, Segment~\footnote{https://segment.com/} and InVision~\footnote{https://www.invisionapp.com/}. The five main reasons that switch back to monolith are: cost, complexity, scalability, performance and organization. During the process, there are six key aspects needed to be aware of: (1) stop developing more services, (2) consolidate and test paths, (3) unify data storage, (4) implement message bus principle, (5) give up diverse techniques and (6) learn to use modular design principles. Opinions of other practitioners are mixed, but most still believe that the decision to switch back to monolith requires careful consideration of the actual system situation and principles.

\section{Methodology}

Here we aim to understand the state of the arts regarding the methods, techniques, and tools facilitating the shift from microservice back to the monolith, as well as the practitioners' opinions and advice towards such practices. To such an end, we conducted a multivocal literature review (MLR) based on the guidelines defined by ~\cite{garousi2019guidelines}. An MLR is a combination of two parts, including 1) a Systematic Literature Review (SLR) on the academic literature (white) published in journals or conferences, and 2) that on the grey literature, e.g., blog posts, social media posts, and videos~\cite{garousi2019guidelines}. Herein, we used the search query \textit{(microservice* OR micro-service* OR "micro service*") AND monolith* AND (back OR return* OR refactor* OR rearchitect* OR migrat* OR revert* OR re-architect*)} in both white and grey literature search. From the search results, we aimed to select the articles (white and grey) that propose \textit{change back from microservice to monolith} and provide factual evidence and/or practical advice related to real industrial cases. By following the traditional SLR process, we obtained only one academic paper from four sources (Scopus, IEEE, ACM and Web of Science). Furthermore, by searching on Google, Reddit, Quora and Stack Overflow, we obtained 19 useful articles and 9 extra from snowballing~\footnote{All selected articles are listed in Appendix which is saved in Arxiv.org. The link will be shared when the paper is accepted.}. We extracted the data to answer the RQs by adapting and merging the categories provided by the selected articles.

\section{Results}
\label{sec:examples}
According to the review, there are four cases that switch from microservice back to monolith: Istio~\ref{SA2}, Amazon Prime Video~\ref{SA5}, Segment~\ref{SA14} and InVision~\ref{SA20}. Among them, the case of Amazon Prime Video is the most discussed with nine articles. These sources were first published in 2018 and did not attract much attention at first. With the case of Segment in 2020, some discussion among practitioners was generated. The case of Amazon Prime Video in 2023 brought the heated discussion of switching back to monolith.

\subsection{RQ1: What are the reasons for switching back to monolith?}

From the review, we identified five main reasons that cases switch back to monolith.


\textbf{Cost}. Cost is the most common reason why companies switch from microservice back to monolith. In Istio, marginal costs and operational costs are high due to the microservice architecture~\ref{SA1} ~\ref{SA2} ~\ref{SA3} ~\ref{SA4}. Amazon Prime Video has the most serious cost problem in four cases. It uses of serverless components resulted in the overall cost of all the building blocks not allowing for the large-scale acceptance of the solution, and the way the video frames (images) are passed between the different components is expensive for the large number of Tier-1 calls to S3 buckets.~\ref{SA5} ~\ref{SA6} ~\ref{SA7} ~\ref{SA8} ~\ref{SA9} ~\ref{SA10} ~\ref{SA11} ~\ref{SA12} ~\ref{SA13}. The cost-benefit of Prime Video's switch back to monolith is also the most significant. According to official reports, moving the service from microservice back to a monolith reduced the infrastructure cost by over 90\%~\ref{SA5}. Segment also has a cost problem. The operational costs of supporting microservices are unaffordable for it~\ref{SA15} ~\ref{SA18}. Finally, in the case of InVision, it makes the comment that " Microservices Also Have a Dollars-And-Cents Cost"~\ref{SA20} ~\ref{SA21}. The service runs on the server, talks to the database, reports on metrics, and generates log entries, all of which have a very real dollar and cent cost. Therefore, it is evident that microservices incur expenses, particularly when accounting for the necessity of maintaining redundancy for establishing a highly available system~\ref{SA20}.


\textbf{Complexity}. Complexity is one of the most important reasons why companies switch from microservice back to monolith. In the case of Istio, microservice architecture leads to greater complexity. Firstly, different planes in Istio are written in different programming languages~\ref{SA2}. Secondly, different teams are responsible for different services separately, but the reality is that this approach makes for increased complexity and has a bad impact on user usability, rather than making it simpler for the development team to manage~\ref{SA2} ~\ref{SA4}. In addition, all components in Istio's control plane are always released in the same version at the same time, while the functionality of the decoupled version of microservices complicates it~\ref{SA2}. Finally, Istio has only limited isolation, making full isolation of microservices difficult~\ref{SA1} ~\ref{SA2} ~\ref{SA3}. However, some other factors lead to greater complexity in the case Segment except the nature of the microservice architecture itself: managing multiple repositories and divergence of shared libraries~\ref{SA14} ~\ref{SA16} ~\ref{SA19}. Initially, each destination was divided into separate services but the same repository, but this caused frustration and inefficiency. A single broken test affected all destinations, and deploying changes required fixing unrelated tests~\ref{SA14} ~\ref{SA16} ~\ref{SA19}. Breaking out the code for each destination into separate repositories increased complexity and maintenance effort. To ease the burden of developing and maintaining these codebases, it created shared libraries to make common transforms and functionality~\ref{SA14}. The complexity of the problem InVision encountered is very similar to Segment. As time went on, InVision had more repositories, more programming languages, more databases, more monitoring dashboards, etc. that became too much for the development team to bear~\ref{SA20} ~\ref{SA21}.


\textbf{Scalability}. The advantage of microservice scalability becomes a disadvantage in these cases. The control plane costs in Istio are mainly determined by the individual feature (XDS)~\ref{SA1} ~\ref{SA2} ~\ref{SA3}. In contrast, all other functions have marginal costs and the value of isolation is very small~\ref{SA2} ~\ref{SA4}. Amazon Prime Video met a scaling bottleneck. Due to the microservices architecture, it hit a hard scaling limit with around 5\% of the expected load, resulting from the orchestration management implemented using AWS Step Functions~\ref{SA5} ~\ref{SA6} ~\ref{SA7} ~\ref{SA10} ~\ref{SA11} ~\ref{SA13}. Different from Prime Video, Segment scaling challenges are the reason for the automated scaling configuration. As the number of destinations grows, managing and scaling each microservice becomes a significant operational overhead~\ref{SA14}. Each service has a specific load pattern that requires manual scaling to cope with unexpected spikes~\ref{SA17}. Tuning the auto-scaling configuration becomes more challenging due to the different resource requirements of each service~\ref{SA14} ~\ref{SA18}.


\textbf{Performance}. Performance is the most important reason for Segment's switch back to monolith. The most serious is the head of line blocking. The microservice architecture causes head-of-line blocking, causes delays in all destinations~\ref{SA14} ~\ref{SA16}. This affects the timeliness of event delivery and also customer satisfaction~\ref{SA17}. In addition, high complexity is also a factor that causes system performance to decline. The high complexity caused by microservice puts development teams in a difficult situation where the benefits of modularity and autonomy become burdensome, slowing them down and reducing productivity, which leads to poorer performance~\ref{SA14}.


\textbf{Organization}. The suboptimal management of teams is also a headache, especially in Istio and InVision. In the case of Istio, although microservices allow different teams to manage services individually, in practice this creates a mess for development teams who want simpler management~\ref{SA2}.In contrast, InVision has the most serious people problem. InVision had a legacy team with fewer people but more repositories, databases, programming languages, etc~\ref{SA20}. As time went on, the benefits of Conway's Law became a burden on the legacy team because of this unsuitable 'size', so it was necessary to merge microservices back into a monolith~\ref{SA20} ~\ref{SA21}.

\subsection{RQ2: What are the key aspects to pay attention to during the switching back?}
We analyzed six key aspects during the process that switching from microservice back to monolith.


\textbf{Stop developing more services}. This means new services cannot be introduced. Switching from microservice back to monolith requires an existing microservice to be used as the "center" of the future monolith to host the new functionality~\ref{SA22}. All other services will eventually be merged into this center. However, if we still make new services after switching back to monolith, this could lead to the whole system getting messy.


\textbf{Consolidate and test paths}. Microservices sometimes have a single, coherent flow between multiple systems. Consolidation paths are necessary when systems are merged from microservices to a monolith~\ref{SA22}. After the merger, it is also important to test the path to ensure that the new system runs smoothly. This process ensures that the new monolithic architecture works properly and meets all requirements~\ref{SA23}.


\textbf{Unify data storage}. Shanea proposed there are two main options: move the data to a single database or keep the data separate~\ref{SA23}. The former can reduce costs and improve performance while reducing the complexity of the system. The latter can help maintain the autonomy and isolation of separate components, while still moving towards a more homogeneous structure~\ref{SA23}. The choice of data storage is critical and development teams need to choose carefully based on actual requirements.


\textbf{Implement the message bus principle}. Implementing a message bus, like Kafka, can be a layer of indirection while transitioning~\ref{SA23}. This strategy enables a gradual consolidation of microservice into monolith without any interruptions to the current system. By utilizing a message bus, smooth communication between various components is ensured, facilitating the decomposition and recombination of services as required.


\textbf{Give up diverse techniques}. The feature of microservice is that different services can use different languages, frameworks, etc. And after the system has switched back to monolith, these diversifications need to be given up. For example, most systems should use no more than two back-end languages at any one time~\ref{SA22}.


\textbf{Learn to use modular design principles}. We need to maintain a modular design when switching back to monolith. Modular design allows the code to be organized into distinct modules with clear boundaries, which promotes separation of concerns and maintainability~\ref{SA22}. The modular design also allows systems to gain the flexibility and modularity benefits of microservices with the simplicity and ease of use of monoliths~\ref{SA23}.

\subsection{RQ3: What are the opinions of the other practitioners regarding such "switch-back"?}

Other practitioners have mixed opinions about this 'switch-back' behavior. Some argue that this way is correct. They think microservice is not the “utopian application architecture”~\ref{SA3}. David Heinemeier Hansson scoffs microservice is a zombie architecture~\ref{SA7}. Monoliths do have an advantage over microservices because they are easier to code, scale, deploy, test, and deal with cross-domain issues~\ref{SA24} ~\ref{SA29}. Some still don't agree with it. They believe that microservices are still one of the most popular architectures. Angel posts monoliths are not the solution, and organizations need to think better and support proactively communication channels that can supply the gaps between teams~\ref{SA25}. However, most practitioners still believe that the need to switch back to a monolithic architecture requires consideration of the actual system situation and principles. Such a switch back would require an assessment of whether monolithic is really the best fit for the company's team size, structure, skills, and operational capabilities~\ref{SA23} ~\ref{SA27}. Moreover, most of the disadvantages of microservices are well-known~\ref{SA26}, so Itiel believes that the recommendation for architecture depends on the type of project~\ref{SA28}.

\section{Conclusions}
There are discussions among practitioners regarding switching from microservice back to monolith when, especially, some companies have already taken action. Though it is still too early to claim it as a trend, the practitioners' opinions are certainly worth noticing. In this work, we performed a preliminary investigation on the reasons for companies to decide to switch back to monolith and key aspects to pay attention to during the process. At the same time, we analyzed the opinions of other practitioners regarding this trend. By systematically addressing 29 white and grey literature in the field, our findings reveal cost is the most important reason why companies switch from microservice back to monolith. Furthermore, complexity, scalability, performance and organization are also the main reasons for this trend. During this process of switching back, we summarized there are six key aspects that needed to pay attention to: (1) stop developing more services, (2) consolidate and test paths, (3) unify data storage, (4) implement message bus principle, (5) give up diverse techniques and (6) learn to use modular design principles. The results show that practitioners have started seriously considering the benefits and motivation of switching back. Though academic studies and industrial applications of microservices are obviously in their prime, the pains of microservices and the benefits of adopting monolith can still provide insights into their improvement. In future studies, we shall further investigate the in-depth opinions of the industry on this topic via surveys and interviews. We shall also conduct comparative case studies on the performances of microservice and reversed monolith systems.



\bibliographystyle{plain}
\bibliography{main.bib}

\newpage

\input{AppendixA}

\end{document}

%% file: AppendixA.tex
\section*{Appendix A: The Selected Articles (SAs)} 
\label{The Selected Papers}

{\small
  \begin{enumerate}[labelindent=-5pt,label={[SA}{\arabic*]}]
\item \label{SA1}
Nabor C Mendonça, Craig Box, Costin Manolache, and Louis Ryan. The monolith strikes back: Why istio migrated from microservices to a monolithic architecture. IEEE Software, 38(05):17–22, 2021.

\item \label{SA2}
Craig Box. Introducing istiod: simplifying the control plane. https://istio.io/latest/blog/2020/istiod/, 2020.

\item \label{SA3}
Christian Posta. Istio as an example of when not to do microservices. https://blog.christianposta.com/ microservices/istio-as-an-example-of-when-not-to-do-microservices/, 2020.

\item \label{SA4}
Luca Bianchi. Istio is moving back to the monolith, but it doesn’t mean you have to do the same. https:// aletheia.medium.com/istio-back-to-monolith-and-you-88dd3bd23265, 2020.

\item \label{SA5}
Marcin Kolny. Scaling up the prime video audio/video monitoring service and reducing costs by 90\%. https://www.primevideotech.com/video-streaming/scaling-up-the-prime-video-audio-videomonitoring -service-and-reducing-costs-by-90, 2023.

\item \label{SA6}
Joab Jackson. Return of the monolith: Amazon dumps microservices for video monitoring. https://thenew stack.io/return-of-the-monolith-amazon-dumps-microservices-for-video-monitoring/, 2023.

\item \label{SA7}
Richi Jennings. Microservices sucks — amazon goes back to basics. https://devops.com/microservices-amazon-monolithic-richixbw/, 2023.

\item \label{SA8}
Tim Anderson. Reduce costs by 90\% by moving from microservices to monolith: Amazon internal case study raises eyebrows. https://devclass.com/2023/05/05/reduce-costs-by-90-by-moving-from-microservi ces-to-monolith-amazon-internal-case-study-raises-eyebrows/, 2023.

\item \label{SA9}
Ed Targett. Amazon prime video team throws aws serverless under a bus. https://www.thestack.technology /amazon-prime-video-microservices-monolith/, 2023.

\item \label{SA10}
Craig Jellick. Microservices are dead, long live the monolith. https://www.acorn.io/microservices-are-dead-long-live-the-monolith/, 2023.

\item \label{SA11}
Neal Weinberg. 6 lessons from the amazon prime video serverless vs. monolith flap. https://www.network world.com/article/3697737/6-lessons-from-the-amazon-prime-video-serverless-vs-monolith-flap.html , 2023.

\item \label{SA12}
David Mooter. The death of microservices? https://www.forrester.com/blogs/the-death-of-microservices, 2023.

\item \label{SA13}
Anshita Bhasin. Exploring amazon prime video’s architecture: Migrating from microservices to monolith for audio/video monitoring service. https://medium.com/@anshita.bhasin/exploring-amazon-prime-vid eos-architecture-migrating-from-microservices-to-monolith-for-aacbf9fabc73, 2023.

\item \label{SA14}
Alexandra Noonan. Goodbye microservices: From 100s of problem children to 1 superstar. https://segme nt.com/blog/goodbye-microservices/, 2018.

\item \label{SA15}
Arun Ramakani. It’s back! a trend to build monolith. https://itnext.io/its-back-a-trend-to-build-monolith-852aaa5e086f, 2020.

\item \label{SA16}
Thomas Betts. To microservices and back again - why segment went back to a monolith. https://www.infoq .com/news/2020/04/microservices-back-again/, 2020.

\item \label{SA17}
Dan Meyer. Segment struggled with microservices, went back to monolith. https://www.sdxcentral.com/ articles/news/segment-struggled-with-microservices-went-back-to-monolith/2018/08/, 2018.

\item \label{SA18}
Adfolks. Transitioning from monolithic to microservices architecture: Pros, cons, and segment’s journey. https://www.adfolks.com/blogs/transitioning-from-monolithic-to-microservices-architecture-pros-cons-and-segments-journey, 2021.

\item \label{SA19}
Alexandra Noonan. To microservices and back again. https://www.infoq.com/presentations/microservice s-monolith-antipatterns/, 2020.

\item \label{SA20}
Ben Nadel. Why i’ve been merging microservices back into the monolith at invision. https://www.bennade l.com/blog/3944-why-ive-been-merging-microservices-back-into-the-monolith-at-invision.htm, 2020.

\item \label{SA21}
ForkMyBrain. Why i’ve been merging microservices back into the monolith at invision. https://notes.nicol evanderhoeven.com/readwise/Articles/Why+I’ve+Been+Merging+Microservices+Back+IntotheMonolith+ at+InVision, 2020.

\item \label{SA22}
David Heinemeier. How to recover from microservices. https://world.hey.com/dhh/how-to-recover-from-microservices-ce3803cc, 2023.

\item \label{SA23}
Shanea Leven. The great "un-migration": Migrating from microservices back to a monolith. https://learn .co desee.io/migrating-from-microservices-to-a-monolith/, 2023.

\item \label{SA24}
Fernando Doglio. Microservices aren’t always the answer: a case for monoliths. https://blog.bitsrc.io/when -microservices-are-not-the-answer-a-case-for-monoliths-895ccefa2728, 2022.

\item \label{SA25}
Angel Paredes. Bring back the monolith. https://medium.com/glovo-engineering/bring-back-the-mono lith-92de928ae322, 2022.

\item \label{SA26}
Diego De Sogos. Microservices’s dark side: The monolith strikes back. https://www.hexacta.com/microse rvicess-dark-side-the-monolith-strikes-back/, 2020.

\item \label{SA27}
Ciaran O’Donnell. Monolith to microservices, and back again? https://ciaranodonnell.dev/posts/the-microservices-monolith-pendulum/, 2020.

\item \label{SA28}
Itiel Maayan. Will modular monolith replace microservices architecture? https://medium.com/att-israel/ will-modular-monolith-replace-microservices-architecture-a8356674e2ea, 2022.

\item \label{SA29}
Shai Almog. Is it time to go back to the monolith? https://dev.to/codenameone/is-it-time-to-go-back-to-the-monolith-3eok, 2023.

\end{enumerate}